\documentstyle[12pt]{article}
\begin{document}
\thispagestyle{empty}

\centerline{\large\bf CHL Compactifications and Beyond}

\bigskip
\centerline{\bf Shyamoli Chaudhuri\footnote{Email: shyamolic@yahoo.com}}
\medskip
\centerline{1312 Oak Drive}
\centerline{Blacksburg, VA 24060}
\bigskip
\begin{abstract}
This is the transcript of a talk given at the {\em 3rd Simons Workshop in 
Mathematics and Physics} on July 26, 2005. We review the genesis of the 
CHL (Chaudhuri-Hockney-Lykken) project, explaining both its phenomenological 
goals and theoretical 
justification in light of the known vast proliferation of N=1 string vacua.
We explain what a CHL compactification
is, review some key results such as the construction of
moduli spaces with a {\em small} number of massless scalar fields, the 
decompactification of such moduli spaces to one of five consistent 
ten-dimensional superstring theories, and the
appearance of electric-magnetic duality in only the {\em four-dimensional}
moduli spaces, a 1995 observation due to Chaudhuri \& Polchinski. 
\end{abstract}

\bigskip\bigskip
\vskip 0.1in One of the goals of this workshop is to determine
the precise boundaries of String Theory, thereby discovering the 
principles by which we can eliminate the redundancy of the 
multiplicity of string vacua and arrive at a convincing description of the
real world. This has been the theme of my research for many years, and
it evolved from the 1995 discovery of the CHL Strings
\cite{chl,cp}.
I therefore decided to describe this
discovery in my talk today, highlighting some of the results and 
follow-up insights I have found 
especially significant \cite{land}.

\vskip 0.1in The CHL compactifications are supersymmetry 
preserving orbifolds of any consistent compactification of
perturbative superstring theory, where by consistency we mean here
an exactly solvable background of perturbative string theory 
to all orders in the $\alpha^{\prime}$ expansion. Such backgrounds 
have a solvable superconformal field theory description on the worldsheet, 
leading to an anomaly-free, and ultraviolet finite, perturbatively 
renormalizable
superstring theory in target spacetime. We use the term {\em perturbatively 
renormalizable}  to describe such a target spacetime string theory Lagrangian 
despite the presence of  infinitely many couplings in the $\alpha^{\prime}$ expansion, 
because {\em only a finite number of independent parameters} go into their determination, 
and these can all be found at the lowest orders in the string effective Lagrangian.
The existence of only a finite number of independently renormalized couplings
is the defining criterion for the Wilsonian renormalizability of a quantum theory.
Thus, from this 
perspective, the perturbative string theoretic unification of gravity and Yang-Mills gauge
theories with chiral matter can be seen as providing a precise, and unique, 
gravitational extension of the 
anomaly-free and renormalizable Standard Model of Particle Physics. 

\vskip 0.1in Of course, since we lack a precise formulation of 
nonperturbative string theory at the
current time, we can only reliably invoke the above 
framework as long as we remain within the domain of weak string coupling.
Fortunately, all observational signals point to the weak unification of the gauge
couplings and gravity in our four-dimensional world, with the supersymmetry breaking 
scale lying somewhere between the electroweak (TeV) and gauge coupling unification 
($10^{16-17}{\rm GeV})$ scales. So perhaps we will be lucky and able to
follow the target spacetime string effective Lagrangian approach up to at 
least the gauge coupling unification scale, using tried-and-true renormalization group methods.
Indeed the stream of precision data from the Z factories in the early 90's, pinning down 
both the number of lepton-quark generations, as well as the
hierarchical texture of fermion masses with the discovery of a surprisingly heavy top quark, 
and increasingly tight windows on neutrino masses, stimulated the resurgence of
theoretical investigations of supersymmetric grand unification models using RG techniques. 

\vskip 0.1in It was in light of
these developments that Joe Lykken and I initiated an ambitious new effort in string model
building in 1993-94. Our original goal was to identify
exactly solvable conformal field theory (cft) realizations of four-dimensional heterotic string vacua 
with massless particle spectra and couplings that would cover the spread of plausible
semi-realistic extensions of the supersymmetric Standard Model, perhaps even suggest
some new features unforseen in conventional, field theoretic model building. We focussed
on fermionic current algebra realizations of the conformal field theory, using a formalism originally
developed
by Kawai, Lewellen, Schwartz, and Tye, because of the 
simple, and explicit, nature of this description. It is straightforward to {\em embed} the
desired particle spectrum and couplings in the cft using the fermionic representation;
the symmetries of the low energy string 
spacetime effective Lagrangian are transparent. Each of these exact cft solutions correspond to 
special points in the moduli space of some CHL orbifold with a chiral, 4D N=1 supersymmetry. 
I should mention that the detailed checks of the worldsheet constraints in such string
vacua are
prohibitively calculation-intensive.
The necessary algorithm had to be implemented on the computer with extensive software support
from George Hockney, also at Fermilab. Remnant ambiguity in the implementation of 
modular invariance for the fermionic twisted $Z_2$ current algebras was ironed out 
by invoking the Verlinde fusion rules, an analysis by Joe
Lykken, in collaboration with a Fermilab postdoc, Stephen-wei Chung \cite{consist}.

\vskip 0.1in
Once on board, however, Hockney's interactive computer program, {\em Spectrum}, was
beautifully simple to utilize for even sophisticated phenomenological model building. 
Certainly,
this was the stage at which I became quite active in finding 
new, and unexpected, cft solutions.
Let me outline the
phenomenological successes of just one of our 4d N=1 examples, designated 
the CHL5 Model in the literature \cite{chl2,cvet}. It describes an N=1 heterotic
string vacuum
with three generations of supersymmetric Standard Model 
$SU(3)_C$$\times$$SU(2)_L$$\times$$U(1)_Y$ particles, an anomalous
$U(1)$, and a very small number of flat directions at the string scale. Analysis
of the flat directions removes all but one additional $U(1)$ at the string scale
in the anomaly-free vacuum, the hypercharge embedding mimics the $SU(5)$ 
result extremely well, $k_Y$$=$$11/6$, quite close to $5/3$, without actual 
grand unification, giving acceptable values for the gauge coupling unification
scale. One generation is singled out from the other two by its distinct couplings
already at the string scale. The breaking of the additional $U(1)$ at either an 
intermediate, or electroweak, scale can generate interesting fermion mass 
textures. The detailed implications of either scenario for CHL5 
have been explored in the papers by Cleaver, Cvetic, Espinosa, Everett, 
and Langacker \cite{cvet}. 

\vskip 0.1in It should be emphasized that CHL5 is already a rather good
string theory description of the observable sector of supersymmetric standard model 
particle physics. But the hidden sector of this model is rather heavily constrained, and
not terribly interesting. Helpful
new input that would enable one to improve on such model-building exercises 
is expected to come in the near future when LHC turns on, giving
insight into both the supersymmetry breaking scale as well as, hopefully, the mass of the 
lightest supersymmetric partner. Inputs from the crucial neutrino sector are already
the focus of
both current, and future, astro-particle experiments. Perhaps whole 
classes of inflationary models can be ruled out so
that we will have additional insight both into viable mechanisms for supersymmetry
breaking, as well as the viable early Universe scenarios.
I should emphasize that all of this is nothing more than physics
extracted from the {\em leading} terms in the string 
spacetime effective Lagrangian: the infrared limit of the full perturbative
string theory, and in a specific 4D flat target spacetime background. 

\vskip 0.1in 
The nonperturbative, pre-spacetime-geometry framework for string/M theory that 
may lie below the 
distance scale at which target spacetime Lagrangians become
our primary investigational tool is addressed in my recent paper \cite{lands}. 
A rather important question that needs to be addressed at this
juncture 
is the apparent {\em disconnectedness} of the vacuum landscape of String Theory
evidenced by the discovery of the CHL moduli spaces \cite{chl,cp}, a
phenomenon that raises the spectre of both island Universes, and of a
fundamental necessity for the 
Anthropic Principle. 
The theoretical paradigm that will enable us to understand these issues 
is provided by the Hartle-Hawking framework for 
Quantum Cosmology \cite{hawk}, but let me begin by explaining the
nature of the CHL moduli spaces. 

\vskip 0.1in The CHL (Chaudhuri-Hockney-Lykken) strings were
named by Polchinski \cite{cp} for the 
authors of the 1995 paper that pin-pointed the existence of additional exactly solvable
 supersymmetry
preserving solutions to the heterotic string theory consistency 
conditions other than toroidal compactifications \cite{chl}.
As explained above, the original motivation for the study of the CHL compactifications was better 
4d N=1 low energy susy particle phenomenology, and a serious wrinkle on such
efforts had been the proliferation of massless scalar moduli in any semi-realistic 
examples. To understand why the CHL compactifications have many fewer flat
directions, consider the simplest example, the 
 Chaudhuri-Polchinski 
orbifold of the circle compactified 
$E_8$$\times$$E_8$ heterotic string described in \cite{cp}:
\begin{equation}
p = {{1}\over{{\sqrt{2}} }} (p_1 + p_2 )  \in \Gamma_8 , \quad  
p = {{1}\over{{\sqrt{2}} }} (p_1 - p_2 )  \in \Gamma_8^{\prime} , \quad p_3 \in \Gamma^{(1,1)} 
\quad ,
\label{eq:mom}
\end{equation}
Let us mod out by the ${\rm Z}_2$ outer automorphism, $R$, interchanging
the two $E_8$ lattices, $\Gamma_8$$\oplus$$\Gamma_8^{\prime}$,
accompanied by a translation, $T$, in the (17,1)-dimensional momentum lattice,
$({\bf v}, 0 ; {\bf v}_3)$ \cite{cp}. This projects onto the symmetric linear 
combination of the momenta in the two $E_8$ lattices, so that the gauge
group is generically $E_8$$\times$$U(1)$:
\begin{equation}
p = {{1}\over{{\sqrt{2}} }} (p_1 + p_2 )  \in \Gamma_8 , \quad  
p = {{1}\over{{\sqrt{2}} }} (p_1 - p_2 )  \in \Gamma_8^{\prime} , \quad p_3 \in \Gamma^{(1,1)} 
\quad ,
\label{eq:mome}
\end{equation}
The RT orbifold acts on the perturbative heterotic string spectrum 
as follows \cite{cp}. Let ${\cal U}$ and ${\cal T}$ denote, respectively,
the untwisted and twisted sectors of the orbifold. The untwisted sector is
composed of states invariant under R (${\cal I}$), and under T (${\cal I}^*$):
\begin{eqnarray}
{\cal I}:&& \quad \quad p_1 = 0 , \quad p_2 = {\sqrt{2}} \Gamma_8 , \quad 
p_3 \in \Gamma^{(10-d,10-d)} 
\cr
{\cal I}^*: &&\quad \quad p_1 = 0 , \quad p_2 = {{1}\over{{\sqrt{2}} }} 
\Gamma_8 , \quad 
p_3 \in \Gamma^{(10-d,10-d)} \cr
{\cal T}: && \quad \quad 
{\bf p} \in {\cal I}^* + {\bf v} \quad . 
\label{eq:orb}
\end{eqnarray}
Notice that the dimension of the moduli space is much {\em smaller}:
the massless scalars parametrize the coset, $SO(18-d,10-d)/SO(18-d)\times SO(10-d)$,
upto discrete identifications, for compactification on the torus $T^{10-d}$. 
The momentum
vectors in the Hilbert space of the orbifold lie on hyperplanes within the 
(26-d,10-d)-dimensional lattice describing the toroidally compactified
$E_8$$\times$$E_8$ string. Such moduli spaces include several novelties
including affine Lie algebra realizations of the simply-laced gauge groups
at higher Kac-Moody level, as well as enhanced symmetry points with non
simply laced gauge symmetry \cite{chl,cp}. We will return to these novelties below.

\vskip 0.1in 
Notice that since the orbifold
action in question {\em preserves} supersymmetry, our discussion of the disconnectedness of
the CHL moduli spaces with 16 supersymmetries can be carried over to 
an analogous disconnectedness of CHL moduli spaces constructed 
as orbifolds of 4D heterotic vacua with 12, 8, 4, or even zero, supersymmetries. 
Consider  the decompactification limit to ten dimensions 
without any change in the number of supersymmetries: for the $Z_2$ orbifold, 
the outer automorphism interchanging the two $E_8$ lattices
becomes trivial in this limit, and we straightforwardly recover an additional $248$
massless gauge bosons. A subsequent toroidal compactification completes the 
continuous interpolating path connecting a point in the CHL moduli space with some point
in the moduli space of toroidal compactifications, via the ten-dimensional 
$E_8$$\times$$E_8$ heterotic string ground state. Thus, a spontaneous decompactification to
ten dimensions followed 
by a spontaneous re-compactification can indeed interpolate
between a pair of (dis)connected CHL moduli spaces with different numbers of abelian 
multiplets. But this is a genuinely stringy phenomenon, with a slew of
modes with string scale masses descending into the massless field theory as we tune
the compactification radius to its noncompact limit.

\vskip 0.1in A similar example is the spontaneous restoration of extended supersymmetry
known to occur in certain decompactification limits of the moduli space of 
4d $N$ $=$ $2$ heterotic string compactifications \cite{koun1}. The modular invariant one
loop vacuum amplitude of the freely acting orbifold in question is parameterized by
the continuously varying, complex structure moduli and Kahler moduli of a six-torus, in addition 
to a constant background electromagnetic field; the
extended supersymmetry is restored in the limit that one of the cycles of the torus
decompactifies \cite{koun1}. Does this framework allow for continuous
interpolations between the moduli space of toroidal compactifications of the 
heterotic string and a CHL moduli space with eight fewer abelian multiplets along
a path that traverses a family of ground states with only eight supercharges, 
and in one lower spacetime dimension? 

\vskip 0.1in The problem with either of these proposals is that the interpolating 
trajectories are exactly marginal flows from the perspective of the 2d worldsheet
renormalization group. Thus, there is no reason to expect the stringy ground state to
\lq\lq evolve" along such a trajectory in the absence of supersymmetry breaking with
a consequent lifting of the vacuum degeneracy. In other words, if the 
supersymmetry breaking scale in Nature
does turn out to be significantly lower than the string scale, the stringy massive 
modes in the CHL moduli space will have genuinely decoupled from the low energy 
field theory limit, and there is no escaping the conclusion that the field theoretic dynamics 
of vacuum selection occurs in one of a multitude of disconnected, low energy Universes. 
How should we
interpret the resulting multitude of low energy string 
effective Lagrangians? The Hawking-Hartle paradigm \cite{hawk} would
identify each such low-energy spacetime 
effective Lagrangian as the final state of a {\em consistent 
history} in some putative Quantum Theory of the Universe. The pre-spacetime
matrix framework for nonperturbative String/M theory described in my recent work 
\cite{land} is such a theory, yielding also a multitude of acceptable 
spacetime effective Lagrangians, each characterized by a distinct large N limit 
of the matrix
Lagrangian. The \lq\lq theory" for the
Initial Conditions of the Universe \cite{hawk}, to borrow a phrase from 
Hartle and Hawking, 
is the pre-spacetime finite N matrix dynamics. This dynamics is 
beyond the direct purview of perturbative string theory.

\vskip 0.1in To summarize, if the supersymmetry-breaking scale 
is clearly separated from the string mass-scale, the stringy massive modes will 
have genuinely decoupled from the effective Lagrangian of relevance and there 
is no escaping the conclusion that vacuum 
selection in perturbative string
theory involves more than just 
dynamics, requiring a discrete choice among disconnected low energy Universes. 
However, upon including the behavior of the 
stringy massive modes, 
all of the CHL moduli spaces are indeed connected in the sense that they
decompactify to the same ten-dimensional perturbative string vacuum. 

\vskip 0.1in
By now, the CHL moduli spaces have given many fundamental new insights into 
weak-strong electric-magnetic dualities in the String/M theory web \cite{chl,flux,land}.
Let me mention the earliest of these discoveries which appears in the paper
\cite{cp}; this particular observation is due to Joe Polchinski. Careful examination of
which non-simply laced gauge groups can appear at the enhanced symmetry 
points in the CHL moduli spaces reveals the result \cite{cp}:
\begin{equation}
{\rm Sp} (20-2n) \times SO(17-2d+2n) \quad n=0, \cdots , 10-d 
\quad ,
\label{eq:esp}
\end{equation}
at special points within the same d-dimensional moduli space. Remarkably, the
electric and magnetic dual groups, ${\rm Sp}(2k)$ and $SO(2k+1)$ for given $k$, 
only appear together in the four-dimensional CHL moduli spaces \cite{cp}. This is 
precisely as required by the S-duality of the 4d N=4 theories, constituting 
independent evidence in favor of it. It should be noted that this property follows
as a consequence of the constraints from modular invariance on the orbifold 
spectrum, the worldsheet constraints responsible for the perturbative renormalizability and 
ultraviolet finiteness of the CHL compactifications. As mentioned above, to the 
best of my knowledge, all of the CHL orbifolds 
described in \cite{chl,cp,cl,land} decompactify to one of the five 10d superstring
theories. A classification of the supersymmetry preserving automorphisms of 
Lorentzian self-dual lattices up to lattices of dimension (6,22) would completely pin down
this important issue, also enabling a classification of the enhanced symmetry
points in each moduli space. This is crucial information necessary for any further
exploration of electric-magnetic duality in the 4D CHL moduli spaces. 

\vskip 0.1in
My work on the abelian symplectic orbifolds of six- and four-dimensional 
toroidally compactified heterotic strings  
with David Lowe in \cite{cl} utilized Nikulin's classification of the automorphisms of
(19,3)-dimensional lattices, namely, the cohomology lattices of the 
classical K3 surfaces. Our analysis proceeds as follows: begin at a point in the
moduli space where the (22,6)-dimensional heterotic momentum lattice decomposes  
as $\Gamma^{(19,3)}$$\oplus$$\Gamma^{(3,3)}$. Given Niemeier's enumerative list of 
self-dual lattices up to dimension 24, one can straightforwardly enumerate a large 
number of CHL orbifolds by invoking Nikulin's classification \cite{cl}. For
instance, the ${\rm Z}_2$ orbifold 
described above readily generalizes to ${\rm Z}_n$ orbifolds with $n$$>$$2$
whenever the (19,3) lattice contains $n$ identical component root-lattices. Modding by
the ${\rm Z}_n$ symmetry under permutation, accompanied by an order-$n$ shift vector 
in the (3,3) torus, gives a ${\rm Z}_n$ CHL orbifold. It is evident that 
a complete classification of the four-dimensional CHL moduli spaces would require 
extending Nikulin's analysis to classification of the symplectic automorphisms of the  
(22,6)-dimensional Lorentzian self-dual lattices. 

\vskip 0.1in
In conclusion, the detailed picture of the
string {\em Landscape} with sixteen supercharges given by the study 
of the CHL compactifications 
has gone a long way towards determining the precise boundaries of the
String/M Duality web. Morevoer, this systematic approach can be
successfully applied to a study of the string 
landscape with 12, 8, 4, or 0 supercharges, and for any of the string
theories, heterotic, type I, or type II, as described in my recent work \cite{land}. 

\vskip 0.1in 
 The discussion during, and after, the talk brought up several
sharp questions that may be answerable within the CHL framework. {\bf Vafa}: 
{\em Are there any 4d N=2 superstring vacua without any additional massless 
scalar fields other than the dilaton?} {\bf Rocek}: {\em Are
there any 4d N=4 heterotic string vacua lacking the six right-moving
abelian gauge fields?} Note that the CHL strings include a 4d N=4 example 
which lacks all 22 left-moving abelian gauge fields present in the generic
toroidal compactification \cite{chl}. This theory has no massless scalar fields
other than the dilaton. Addressed in \cite{land}. 
{\bf Vafa, Rocek, Niewenhuizen, Chaudhuri}: {\em 
What is known about
the 4d N=3 theories coupled to matter? Do all of these theories contain points
in the moduli space with an extended N=4 supersymmetry? Does this always 
require decompactification (a degeneration) of one, or more, cycles of the torus?}
Note the relation to the Ferrara-Kounnas asymmetric orbifolds of the type II 
superstrings described in my recent paper \cite{land}. Their analysis includes 
some N=3 examples that are identifiable CHL compactifications. 

\vskip 0.1in Understanding 
the symmetry principles, and the fundamental
degrees of freedom in terms of which nonperturbative String/M theory 
can be formulated, is an important focus of ongoing research in
theoretical high energy physics. Elucidating the web of heterotic-type I-type II
CHL compactifications preserving sixteen or fewer supercharges 
can play a significant role in guiding such work  
since it determines precise boundaries for what we mean by 
{\em string consistency}.

\vskip 0.2in \noindent {\bf Acknowledgments}: I would like to thank Joe Polchinski, 
Joe Lykken, David Lowe, and also George Hockney, for their collaboration
on the research described here. I thank Cumrun Vafa for the invitation to present this
work. It is a pleasure to acknowledge the organizers and participants of the {\em 3rd Simons
Workshop in Mathematics \& Physics}, including
Dan Kabat, Sanjaye Ramgoolam, Martin Rocek, and Andrew Mc Hugh, 
for a stimulating meeting.

\end{document}